\documentclass[12pt,preprint]{aastex}

\newcommand{\te}{\emph{$t^2$}}

\shorttitle{ C and O Galactic Galactic Gradients I}
\shortauthors{Esteban et al.}

\received{}
\begin{document}

\title{Carbon and Oxygen Galactic Gradients I: Observational Values from \ion{H}{2} Region
Recombination Lines  \footnotemark{}}

\author{C\'esar Esteban}
\affil{Instituto de Astrof{\'\i}sica de Canarias, E-38200, La Laguna, Tenerife, Spain}
\email{cel@ll.iac.es}

\author{Jorge Garc\'{\i}a-Rojas} 
\affil{Instituto de Astrof{\'\i}sica de Canarias, E-38200, La Laguna, Tenerife, Spain}
\email{jogarcia@ll.iac.es}

\author{Manuel Peimbert}
\affil{Instituto de Astronom\'\i a, 
Universidad Nacional Aut\'onoma de M\'exico,
Apdo. Postal 70-264, M\'exico 04510 D.F., Mexico}
\email{peimbert@astroscu.unam.mx}

\author{Antonio Peimbert}
\affil{Instituto de Astronom\'\i a,
Universidad Nacional Aut\'onoma de M\'exico,
Apdo. Postal 70-264, M\'exico 04510 D.F., Mexico}
\email{peimbert@astroscu.unam.mx}

\author{Mar\'{\i}a Teresa Ruiz}
\affil{Departamento de Astronom\'{\i}a, Universidad de Chile,
Casilla Postal 36D, Santiago de Chile, Chile}
\email{mtruiz@das.uchile.cl}

\author{M\'onica Rodr\'{\i}guez}
\affil{Instituto Nacional de Astrof\'\i sica, \'Optica y Electr\'onica, Apartado postal 51, Puebla, Pue, Mexico}
\email{mrodri@inaoep.mx}

\and

\author{Leticia Carigi}
\affil{Instituto de Astronom\'\i a, 
Universidad Nacional Aut\'onoma de M\'exico,  
Apdo. Postal 70-264, M\'exico 04510 D.F., Mexico}
\email{carigi@astroscu.unam.mx}

\begin{abstract}
We present results of deep echelle spectrophotometry of eight Galactic \ion{H}{2} regions located at galactocentric 
distances between 6.3 and 10.4 kpc. The data have 
been taken with the Very Large Telescope (VLT) Ultraviolet Echelle Spectrograph (UVES) in the 3100 to 10360 \AA\ 
range. We have derived C$^{++}$ and O$^{++}$ 
abundances from recombination lines for all the objects, as well as O$^+$ abundances from this kind of lines 
for three of the nebulae. The intensity of recombination lines is almost independent on the assumed electron 
temperature as well as on the possible presence of spatial temperature variations or fluctuations inside the 
nebulae. These data allow the determination of the gas-phase C and O abundance gradients of the Galactic 
disk, of paramount importance for chemical evolution models. This is the first time the C gradient is derived 
from a so large number of \ion{H}{2} regions and for a so wide range of galactocentric distances. 
Abundance gradients are found of the form $\Delta$log(O/H) = $-$0.044$\pm$0.010 dex kpc$^{-1}$, 
$\Delta$log(C/H) = $-$0.103$\pm$0.018 dex kpc$^{-1}$, and $\Delta$log(C/O) = $-$0.058$\pm$0.018 dex kpc$^{-1}$. 

\end{abstract}


\footnotetext[1]{Based on observations collected at the European Southern
Observatory, Chile, proposals number ESO 68.C-0149(A) and ESO 70.C-0008(A)}

\section{Introduction
\label{intro}}
The analysis of intensity ratios of collisionally excited lines (CELs) is the standard 
method for deriving ionic abundances in ionized nebulae. Almost all the available determinations of the  
gas-phase Galactic abundance gradients from \ion{H}{2} regions observations are based on this kind of analysis \citep{sha83, 
aff97, deh00}. The use of CELs has the disadvantage that their intensity strongly depends on the assumed electron 
temperature and on the presence of temperature variations or fluctuations over the observed volume 
of the nebula. The most dramatic effect of temperature fluctuations is that ionic abundances are underestimated if 
these fluctuations are present in the objects but not considered in the abundance calculations \citep{pei67}. 
Recombination lines (RLs) of heavy element ions are very faint and have been difficult to observe in 
nebular spectra. Fortunately, the availability of high efficiency echelle spectrographs in large aperture 
telescopes makes feasible the measurement of RLs for an increasing number of nebular objects. Several autors have 
obtained O$^{++}$/H$^+$ and C$^{++}$/H$^+$ from the intensity of \ion{O}{2} and 
\ion{C}{2} RLs for the brightest \ion{H}{2} regions of the Galaxy (Peimbert et al. 1993; Esteban et al. 1998, 1999a,b, 2004, 
hereafter E04; Tsamis et al. 2003; Garc\'\i a-Rojas et al. 2004, hereafter G04); all of them have found that abundance determinations from RLs are systematicaly larger than those 
obtained using CELs, independently of the ion considered. These differences, as well as other discrepancies in the 
nebular properties derived, can be consistently accounted for assuming a \te\ (mean square temperature variation 
over the observed volume) in the 0.02--0.05 range. 
\citet{est99b} were the first authors who estimated the Galactic C and O abundance gradients from RLs in \ion{H}{2} 
regions, but only with three objects and covering a rather small range of galactocetric distances 
(from 6.4 to 8.4 kpc, assuming the Sun at 8 kpc). This gradient was later revised by \citet{est02} and 
G04, these last authors included abundances for an additional \ion{H}{2} region at 
7.46 kpc (NGC 3576).  
In this letter, we present new values of the C and O gradients obtained from very deep echelle spectra 
of eight Galactic \ion{H}{2} regions located at galactocentric distances ranging from 6.3 to 10.4 kpc (assuming the 
Sun at 8 kpc). Some of our sample nebulae were previously observed by our group with other instrumentation -- 
the Orion nebula \citep{est98}, M~8 \citep{est99a}, and M~17 \citep{est99b}, but they have been re-observed for this project. 
Data of some of the objects included 
in this paper have been already published: NGC 3576 (G04) and Orion nebula (E04).

\section{Observations, Data Reduction, and Line Intensities
\label{obsred}}

The observations were made on 2002 March 11 and 12 and on 2003 March 29, 30, and 31 with the Ultraviolet Visual 
Echelle Spectrograph, \citep{dod00}, at the Very Large Kueyen Telescope in Cerro Paranal Observatory (Chile). 
We used the standard settings in the two arms of the spectrograph, covering the region from 3100 to 10360 \AA\ . 
The instrumental setup and the technical 
aspects 
of the observations were the same in all 
cases and are 
described in detail in G04 and E04. The results included in this Letter correspond to a single slit 
position in each nebula corresponding to the brightest part of the object. 
The atmospheric dispersor corrector was used during the observations in order to avoid the effects of 
differential atmospheric refraction along the slit. The total exposure time was very 
different for each case depending on the surface brightness of the nebulae, ranging from 300 
to 600 seconds for the Orion nebula and from 2700 to 9612 seconds for S~311. In all 
the cases, the total exposure time was achieved adding three or more individual spectra in each spectral zone. 
The slit width was set to 3.0" and the slit length was set to 10" in the blue arm and to 12" in the red arm 
in the cases of the Orion nebula and NGC 3576 and 3.0"$\times$10" in both arms for the rest of the objects. 
The effective spectral resolution at a given wavelength is $\Delta \lambda \sim \lambda / 8800$. The 
spectra were reduced using the IRAF\footnotemark{} echelle reduction package, following the standard procedure. 
The standard star EG~274 was observed for flux calibration in the cases of the Orion nebula and NGC~3576 (both 
observed in 2002). The rest of the objects were observed in 2003 and flux calibrated with the standard stars: 
HD~49798, EG~274, and CD~$-$32~9927. 
Line intensities were measured as described in G04 and E04 with the SPLOT routine of the IRAF 
package. Table~\ref{lines} presents the extinction-corrected intensity of the emission lines relevant to the 
results concerning this Letter, except for Orion nebula and NGC 3576, for which complete line lists and 
abundance data have been published in E04 and G04, respectively. 
The complete line lists of the rest of the objects will be published elsewhere. The reddening coefficient, 
$C(H\beta)$, was determined by fitting iteratively the observed and theoretical Balmer decrement following the 
predictions by \citet{sto95} and is included in Table~\ref{abund}. We have used the reddening law of \citet{sea79} 
for all the objects listed in Table~\ref{lines} except in the cases of M8 and M~16 for which we have used the 
reddening law of \citet{car89} following the prescriptions of \citet{san91} and \citet{chi90}. 
The errors 
associated with the line intensities have been estimated following G04. 

\footnotetext{IRAF is distributed by NOAO, which is operated by AURA,
under cooperative agreement with NSF.}

\section{Results and Discussion
\label{results}}

In Table~\ref{abund}, we present the adopted electron density, $N_e$, and temperature, $T_e$, for each object. 
The adopted $N_e$ corresponds to the mean of the values obtained from [\ion{O}{2}], [\ion{S}{2}], 
[\ion{Cl}{3}], and [\ion{Fe}{3}] --and [\ion{Ar}{4}] in several of the objects--, weighted by their 
corresponding uncertainties. In the case of $T_e$ we have adopted two values, one corresponding to 
the high ionization potential ions, $T_{high}$, which is calculated as the weighted mean of the temperature 
obtained from [\ion{O}{3}], [\ion{S}{3}], and [\ion{Ar}{3}] (only temperatures from [\ion{O}{3}] were 
available for M~16 and M~20), and other to the low ionization potential ones, 
$T_{low}$, which corresponds to the weighted mean of the values obtained for [\ion{O}{2}] and [\ion{N}{2}]. 
In any case, intensity ratios of RLs are almost 
independent on $N_e$ and $T_e$ and they are almost unaffected by the values assumed for 
these parameters. The temperature and density determinations were carried out with the IRAF task TEMDEN 
of the package NEBULAR. More details on the determination of the physical conditions can be found in 
G04 and E04. The adopted values for $N_e$, $T_{high}$, and $T_{low}$ are included 
in Table~\ref{abund}. 

\begin{figure}
\begin{center}
\epsscale{1.0}
\plotone{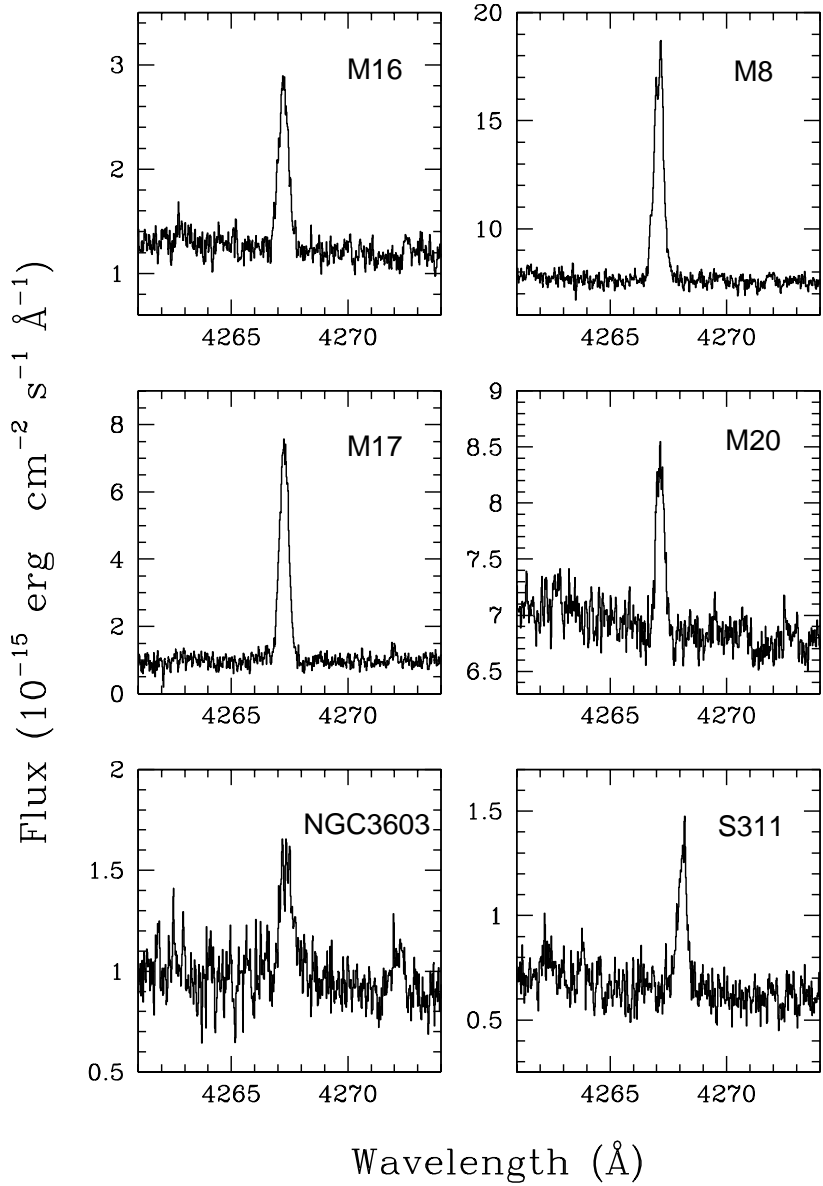}
\figcaption{Section of the echelle spectra of M~16, M~8, M~17, M~20, NGC~3603, and 
S~311 showing the \ion{C}{2} 4267 \AA\ line, the brightest recombination line of 
\ion{C}{2} in the spectral range observed. Similar plots for the Orion nebula and 
NGC~3576 have been published in Esteban et al. (2004) and Garc\'\i a-Rojas et al. (2004), 
respectively. 
\label{cii}}
\end{center}
\end{figure}

We have adopted $T_{high}$ for the derivation of C$^{++}$/H$^+$ and O$^{++}$/H$^+$ ratios and $T_{low}$ 
for O$^+$/H$^+$. We have used the effective recombination coefficients of \citet{sto95} for \ion{H}{1}, 
of \citet{dav00} for \ion{C}{2}, and of \citet{sto94} for \ion{O}{2}. The average of the 
abundances obtained with the effective recombination coefficients of \cite{peq91} and \citet{esc92} has 
been used to obtain the O$^+$/H$^+$ ratio. We have used the intensity of the brightest \ion{C}{2} line 
(\ion{C}{2} 4267 \AA, see Figure~\ref{cii}) to derive the C$^{++}$/H$^+$ ratio. Other much fainter lines of $d-f$ 
and $f-g$ transitions of \ion{C}{2} have been observed in all nebulae except NGC 3603 and S311. These lines give abundances 
in complete agreement with those obtained with \ion{C}{2} 4267 \AA. The only RLs of \ion{O}{1} observed 
in our spectra are those of multiplet 1. However, there are several bright sky lines in that spectral 
region and only \ion{O}{1} 7771.94 \AA\ can be measured. Unfortunately, a satisfactory deblending of the 
\ion{O}{1} line from sky lines was not possible in some of the objects due to the radial velocity of the 
nebulae: Orion nebula, M~16, M~17, and S~311. For these four  nebulae as well as for NGC~3603 (for which 
\ion{O}{1} lines were not detected)  we have determined the O$^+$/H$^+$ ratio from the intensity of 
[\ion{O}{2}] lines using NEBULAR and assuming an appropriate value of $t^2$. The $t^2$ 
adopted for each object is also included in Table~\ref{abund}. This parameter has been determined 
from the weighted mean of the different \te\  values obtained from the comparison of ionic abundances 
obtained from CELs and RLs of the same ions, the comparison of $T_e$ obtained from CELs and from the 
Balmer and Paschen discontinuities, as well as a maximum likelihood method applied to the intensity of 
the brightest \ion{He}{1} lines. Details on the derivation of $t^2$ can be found in G04 and E04. In the 
case of the O$^{++}$ abundances, the values quoted in Table~\ref{abund} are the weighted 
means of the O$^{++}$/H$^+$ ratios obtained from the sum of the intensities of the individual lines of the 
less case-dependent multiplets observed in each nebula following the prescriptions 
given in G04 and E04. We have used  multiplets number 1, 2, 10 and the $4f-3d$ transitions in the cases of 
Orion nebula (E04) and NGC 3576 (G04), multiplets 1, 2, and 10 in the case of M~8 and M~17, and multiplet 1 
in M~16, M~20, NGC~3603, and S~311. The lines of multiplet 1 are much brighter than those of the other 
multiplets and dominate the final mean value, therefore, we have only included the line intensities of 
multiplet 1 in Table~\ref{lines}, the rest of the \ion{O}{2} line intensities of M~8 and M~17 will be 
published elsewhere. 

Once we have the ionic abundances, we can derive the total abundances of the elements. In the case of O, 
we have information of all the relevant ionic species that are expected to be present in the ionized nebulae: 
O$^+$ and O$^{++}$. The total O abundance will be simply the sum of the 
O$^+$/H$^+$ and O$^{++}$/H$^+$ ratios. In the case of C we have only information about the C$^{++}$/H$^+$ ratio. 
Since we do not expect a significant presence of C$^{3+}$ in the sample (see Garnett et al. 1999), the only 
important contribution is of C$^+$, which does not show emission lines in the spectral range observed. 
To obtain the final C/H abundance we have used the ionization correction factor (ICF) for the presence 
of C$^+$ derived by \citet{gar99} from photoionization models (see their Figure 2). This ICF indicates that 
the C$^+$/C$^{++}$ ratio goes from 0.08$\pm$0.04 in the case of NGC~3603 to 2.2$\pm$0.8 in the case of M20, 
the two nebulae with the extreme values of excitation.  
 
\begin{figure}
\begin{center}
\epsscale{1.0}
\plotone{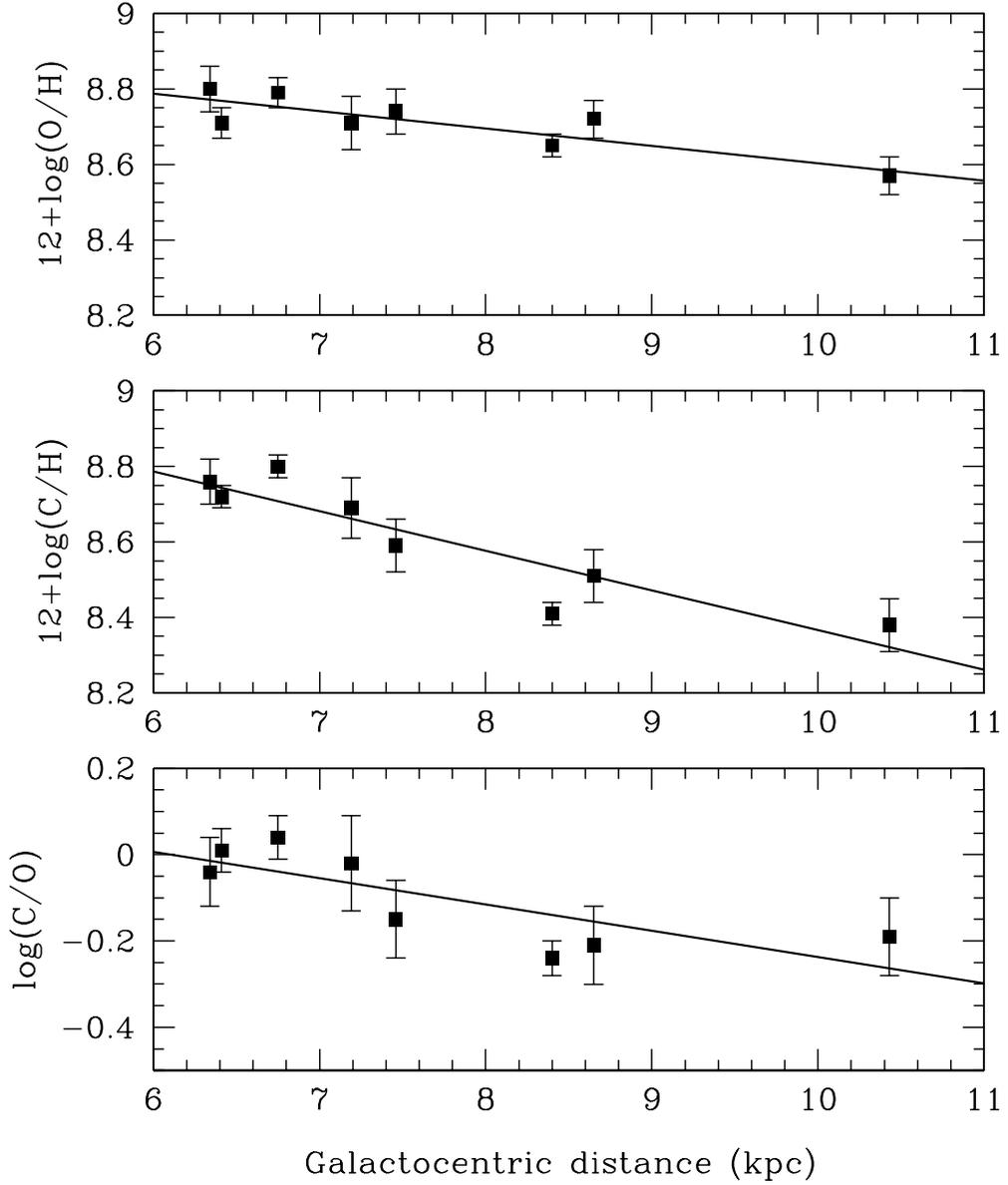}
\figcaption{Ionized gas phase O, C, and C/O radial abundance gradients of the Galactic disk from \ion{H}{2} 
region abundances determined from recombination lines. The lines indicate the least-squares linear fits to 
the data. The Sun is located at 8 kpc. 
\label{grad}}
\end{center}
\end{figure}

 Figure~\ref{grad} 
shows the C/H, O/H, and C/O ratios derived for the eight nebulae of the sample. The galactocentric distances of 
the objects have been taken from the stellar photometric distances quoted in the survey of Galactic star-forming 
complexes of \citet{rus03} and the work on distances to open clusters of \citet{dia02}. The Sun is assumed 
to be at a galactocentric distance of 8 kpc. Simple least-squares linear fits to the data given in 
Figure~\ref{grad} give the following gradients:

12 + log(O/H) = (9.043$\pm$0.082)$-$(0.044$\pm$0.010)$R_G$  (r = 0.647),

12 + log(C/H) = (9.396$\pm$0.139)$-$(0.103$\pm$0.018)$R_G$  (r = 0.905),

log(C/O) = (0.360$\pm$0.141)$-$(0.058$\pm$0.018)$R_G$  (r = 0.762).

These gradients and the data used for their derivation correspond to the composition of the ionized gas-phase of 
the interstellar medium (ISM). 
\citet{est98} estimated that C and O abundances of the Orion nebula should be increased about 0.1 dex and 
0.08 dex, respectively, to obtain the current gas+dust composition of the local ISM. We do not know if these 
depletion factors for the Orion nebula can be applied to all the sample objects 
or if they depend on the particular ionization conditions of the nebulae. In any case, the relatively small 
dispersion of the C and O abundances for a given distance suggests that the possible differences in the 
depletion factors should not be large. Therefore, in the absence of further information, we recommend to 
apply the depletion factors estimated by \citet{est98} to all the objects prior to compare our abundances  
with the gas-phase ones predicted by chemical evolution models. 

The value of the O gradient we derive is shallower than those obtained by \citet{sha83} ($-$0.080 dex kpc$^{-1}$) 
and \citet{aff97} ($-$0.070 dex kpc$^{-1}$) but similar to the value of $-$0.040 dex kpc$^{-1}$ obtained 
by \citet{deh00} and that estimated by \citet{pil03} ($-$0.051 dex kpc$^{-1}$). All these determinations of the 
literature are based on the standard analysis based on the intensity of CELs, therefore, their result depend very 
much on 
the exact value of the $T_e$ assumed for each object. In our case, the O gradient we obtain from CELs and assuming 
$t^2$=0 is found of the form $\Delta$log(O/H) = $-$0.040$\pm$0.006 dex kpc$^{-1}$, which is very similar to the 
gradient we derive from RLs. The Ne, Ar and  and S gradients -which are expected to follow the 
behavior of O- determined by 
Mart\'{\i}n-Hern\'andez,  van der Hulst,  \& Tielens (2003, and references therein) from ISO data are also 
consistent with shallower values of the slope. Our C gradient is clearly much steeper than that we obtain for O. 
Its slope is also steeper than the one obtained 
by \citet{rol00} for B stars ($-$0.070 dex kpc$^{-1}$). The C/O gradient is an important constraint for chemical 
evolution models and the radial star formation history of the Galactic disk because the nucleosynthetic origin 
of both elements is expected to be quite different. We find that the C/O ratio increases with metallicity. 
Our C/O gradient is similar to that 
obtained by \citet{gar99} for two nearby spiral galaxies: M 101 and NGC 2403 ($-$0.04 and $-$0.05 dex kpc$^{-1}$, 
respectively), from the analysis of CELs in the UV domain and is also similar to that obtained by 
\citet{sma01} for Galactic B stars  ($-$0.050 dex kpc$^{-1}$). 
Models of Galactic chemical evolution to explain the observed gradients are presented by Carigi et al. (2004).
C and O are the most important ``biogenic elements'' and their gradients are paramount, in the study of the Galactic 
habitable zone (e.g. Lineweaver, Fenner \& Gibson 2004 and references therein) as well as to determine the CO to 
H$_2$ conversion factor (e.g. Strong et al. 2004 and references therein).

We are grateful to the referee, Dick Henry, for his careful review of this letter. 
CE and JG would like to thank the members of the Instituto de Astronom\'\i a, UNAM, for their always 
warm hospitality. This work has 
been partially funded by the Spanish Ministerio de Ciencia y Tecnolog\'\i a (MCyT) under project AYA2001-0436. 
MP received partial support from DGAPA UNAM (grant IN114601). MTR received partial support from FONDAP(15010003), 
a Guggenheim Fellowship and Fondecyt(1010404). MR acknowledges support from Mexican CONACYT project J37680-E.
LC's work is supported by CONACyT grant 36904-E.

\clearpage

\begin{deluxetable}{c@{\hspace{5pt}}c@{\hspace{5pt}}cccccc} 
\tabletypesize{\scriptsize}
\tablecaption{Reddening corrected relevant line ratios (F(H$\beta$) = 100).
\label{lines}}
\tablewidth{0pt}
\tablehead{
\colhead{$\lambda_0$ (\AA)} &
\colhead{Ion} & 
\colhead{M~16} &
\colhead{M~8} &
\colhead{M~17} &
\colhead{M~20} &
\colhead{NGC~3603} &
\colhead{S~311}} 
\startdata
3726.03 &[O II] & 155$\pm$6& 151$\pm$5& 46$\pm$2& 147$\pm$6& 40$\pm$2& 165$\pm$7\\
3728.82 &[O II] & 122$\pm$5& 106$\pm$4& 46$\pm$2& 173$\pm$7& 25$\pm$1& 194$\pm$8\\
4267.15 & C II & 0.27$\pm$0.03& 0.22$\pm$0.01& 0.58$\pm$0.04& 0.17$\pm$0.02& 0.33$\pm$0.06& 0.108$\pm$0.013\\
4363.21 & [O III]& 0.19$\pm$0.02& 0.29$\pm$0.01& 0.95$\pm$0.05& 0.15$\pm$0.02& 2.5$\pm$0.1& 0.56$\pm$0.02\\
4638.86 & O II & 0.040$\pm$0.012& 0.034$\pm$0.005& 0.093$\pm$0.018& ...& 0.057:& 0.028$\pm$0.008\\
4641.81 & O II & 0.034$\pm$0.012& 0.043$\pm$0.005& 0.128$\pm$0.020& 0.030:&
0.131$\pm$0.033& 0.028$\pm$0.008\\
4649.13 & O II & 0.056$\pm$0.014& 0.041$\pm$0.005& 0.123$\pm$0.019& 0.036:& 0.172$\pm$0.034& 0.025$\pm$0.008\\
4650.84 & O II & 0.050$\pm$0.014& 0.032$\pm$0.005& 0.100$\pm$0.018& 0.017:& 0.086$\pm$0.029& 0.025$\pm$0.008\\
4661.63 & O II & 0.038$\pm$0.013& 0.036$\pm$0.005& 0.119$\pm$0.019& 0.018:& 0.111$\pm$0.031& 0.024$\pm$0.008\\
4673.73 & O II & ...& ...& 0.022:& ...& ...& ...\\
4676.24 & O II & ...& 0.017$\pm$0.004& 0.044$\pm$0.014& ...& 0.040:& ...\\
4861.33 & H I & 100$\pm$3& 100$\pm$3& 100$\pm$3& 100$\pm$3& 100$\pm$3& 100$\pm$3\\
4958.91 & [O III] & 27.8$\pm$0.9& 33$\pm$1& 114$\pm$3& 19.0$\pm$0.6& 180$\pm$5& 43$\pm$1\\
5006.84 & [O III] & 82$\pm$3& 96$\pm$3& 335$\pm$10& 59$\pm$2& 533$\pm$16& 126$\pm$4\\
7771.94 & O I & 0.013:$^{\rm a}$& 0.029$\pm$0.003& 0.025$\pm$0.005$^{\rm a}$& 0.036$\pm$0.006& ...& ...\\
\enddata
\tablenotetext{a}{Blend with sky emission} 
\end{deluxetable}

\clearpage

\begin{deluxetable}{l@{\hspace{5pt}}cccccccc} 
\rotate
\tabletypesize{\scriptsize}
\tablecaption{Some Parameters, Physical Conditions and Chemical Abundances
\label{abund}}
\tablewidth{0pt}
\tablehead{
\colhead{Parameter} & 
\colhead{M~16} &
\colhead{M~8} &
\colhead{M~17} &
\colhead{M~20} &
\colhead{NGC~3576} &
\colhead{Orion neb.} &
\colhead{NGC~3603} &
\colhead{S~311}} 
\startdata
R$_G$ (kpc)& 6.34$^{\rm a}$& 6.41$^{\rm b}$&  6.75$^{\rm a}$& 7.19$^{\rm a}$& 7.46$^{\rm b}$& 8.40$^{\rm a}$& 
8.65$^{\rm a}$& 10.43$^{\rm b}$\\
C(H$\beta$)& 1.21$\pm$0.06& 0.94$\pm$0.03& 1.17$\pm$0.05& 0.36$\pm$0.04& 1.40$\pm$0.07& 0.76$\pm$0.08& 
2.36$\pm$0.06& 0.64$\pm$0.04\\
N$_{\rm e}$ (cm$^{-3}$)& 1120$\pm$220& 1600$\pm$200& 470$\pm$120& 270$\pm$60& 
2800$\pm$400& 8900$\pm$200& 5150$\pm$750& 310$\pm$80\\
T$_{high}$ (K)& 7650$\pm$250& 8040$\pm$130& 8050$\pm$150& 7800$\pm$300& 8500$\pm$150& 8320$\pm$40& 9050$\pm$200& 9050$\pm$200\\
T$_{low}$ (K)& 8350$\pm$200& 8450$\pm$150& 8870$\pm$300& 8400$\pm$200& 8500$\pm$150& 10000$\pm$400& 11400$\pm$700& 9500$\pm$250\\
$t^2$& 0.036$\pm$0.006& 0.037$\pm$0.004& 0.033$\pm$0.005& 0.036$\pm$0.013& 0.038$\pm$0.009& 0.022$\pm$0.002& 0.040$\pm$0.008& 
0.038$\pm$0.007\\
C$^{++}$/H$^+$ $\times$ 10$^5$& 25$\pm$2& 20$\pm$1& 53$\pm$4& 15$\pm$2& 28$\pm$4& 22$\pm$1& 30$\pm$5& 10$\pm$1\\
12+log(C/H)& 8.76$\pm$0.06& 8.72$\pm$0.03& 8.80$\pm$0.03& 8.69$\pm$0.08& 8.59$\pm$0.07& 8.41$\pm$0.03& 
8.51$\pm$0.07& 8.38$\pm$0.07 \\
O$^{+}$/H$^+$ $\times$ 10$^5$& 44$\pm$10$^{\rm c}$& 34$\pm$5& 10$\pm$2$^{\rm c}$& 42$\pm$7&
13$\pm$3& 14$\pm$4$^{\rm c}$& 4$\pm$1$^{\rm c}$& 25$\pm$5$^{\rm c}$\\
O$^{++}$/H$^+$ $\times$ 10$^5$& 20$\pm$2& 17$\pm$1& 48$\pm$2& 10$\pm$5& 42$\pm$5& 37$\pm$1& 49$\pm$6& 11$\pm$1\\

12+log(O/H)& 8.80$\pm$0.06& 8.71$\pm$0.04& 8.76$\pm$0.04& 8.71$\pm$0.07& 8.74$\pm$0.06& 8.65$\pm$0.03& 
8.72$\pm$0.05& 8.57$\pm$0.05\\
\enddata
\tablenotetext{a}{Dias et al. (2002)}
\tablenotetext{b}{Russeil (2003)}
\tablenotetext{c}{Value determined from [OII] lines and assuming the corresponding $t^2$} 
\end{deluxetable}

\clearpage

\clearpage

\end{document}